\documentclass[12pt,preprint]{aastex}

\newcommand{\be}{\begin{equation}}
\newcommand{\ee}{\end{equation}}
\newcommand{\ba}{\begin{eqnarray}}
\newcommand{\ea}{\end{eqnarray}}

\shorttitle{The Primordial Power Spectrum from WMAP Data}
\shortauthors{Mukherjee \& Wang}

\begin{document}

\title{Model-Independent Reconstruction of the Primordial Power Spectrum
from WMAP Data}
\author{Pia~Mukherjee, Yun~Wang}
\affil{Department of Physics \& Astronomy, Univ. of Oklahoma,
                 440 W Brooks St., Norman, OK 73019;
                 email: pia,wang@nhn.ou.edu}

\begin{abstract}

Reconstructing the shape of the primordial power spectrum in a model
independent way from cosmological data is a useful consistency check 
on what is usually assumed regarding early universe physics.  It is also
 our primary window to unknown physics during the inflationary era.
Using a power-law form for the primordial power spectrum
$P_{in}(k)$
and constraining the scalar spectral index and its running, 
\cite{Peiris03} found that the first year WMAP data
seem to indicate a preferred scale in $P_{in}(k)$.
We use two complementary methods:
the wavelet band powers method of \cite{pia1},
and the top-hat binning method of \cite{Wang99}
to reconstruct $P_{in}(k)$ as a free function
from CMB data alone (WMAP, CBI, and ACBAR), or from CMB data 
together with large scale structure data 
(2dFGRS and PCSZ). The shape of the reconstructed $P_{in}(k)$ is consistent
with scale-invariance, although it allows some indication of a preferred 
scale at $k \sim 0.01$Mpc$^{-1}$.
While consistent with the possible evidence for a running of the 
scalar spectral index found by the
WMAP team, our results 
highlight the need of more stringent and independent constraints
 on cosmological
parameters (the Hubble constant in particular) in order
to more definitively constrain deviations of $P_{in}(k)$ from scale-invariance
without making assumptions about the inflationary model.

\end{abstract}

%\end{document}

%\keywords{Cosmology}

\section{Introduction}

The long anticipated first year WMAP data \citep{Bennett03,Spergel03}
have very interesting implications for the state of cosmology today.
On one hand, WMAP results on cosmological parameters are consistent
with and refine previous constraints from various independent and 
complementary observations.
On the other hand, the data seem to indicate a preferred scale in
the primordial scalar power spectrum $P_{in}(k)$ \citep{Peiris03}, 
with or without complementary large scale structure data, 
though not at high significance. If true,
this would contradict the simple assumption of scale-invariance 
of $P_{in}(k)$ made 
by most researchers in cosmology and the prediction of the 
simplest inflationary models, but would be
consistent with earlier lower significance findings of 
\cite{WangMathews02}, \cite{pia1} (hereafter MW03a), and \cite{pia2}.

With WMAP, CMB data continues to be fully consistent with inflation
\citep{Guth81,KolbTurner90,HuDodelson02,PeeblesRatra03}.
WMAP reveals new evidence for inflation from the anti-correlation 
between CMB temperature and polarization fluctuations near $l$ of 150.
Focus is now shifting towards distinguishing between the different
 inflationary models. The simplest models of inflation predict a
power-law primordial matter power spectrum (for example, 
\cite{Linde83,naturalinf,extendedinf}). Thus some efforts
have been focused on constraining slow roll parameters 
\citep{Liddle92,Leach02,Barger03},
evaluated at a certain epoch during inflation, or 
at the Hubble crossing time of a certain scale usually chosen
 to be at the center of the scales probed by observations. The 
primordial power spectrum that results from slow-roll
 inflation can be computed to high accuracy in terms of these
 parameters. The WMAP team fits to observables
 which can be written as derivatives of the slow roll parameters.
However, there are also viable models of inflation which 
predict primordial power spectra which cannot be parametrized 
by a simple power-law 
(for example, 
\cite{Holman91ab,Linde94,Wang94,Randall96,Adams97,Les97}).
In such models, features in 
$P_{in}(k)$ can result
 from unusual physics during inflation
\citep{Chung00,Enqvist00,Lyth02}.
The assumption of a power-law $P_{in}(k)$ could then lead to
our missing the discovery of the possible features
in the primordial matter power spectrum 
and erroneous estimates of cosmological parameters
\citep{Kinney01}.

WMAP results underscore the importance of model-independent
measurements of the shape of the primordial power spectrum \citep{Wang99}.
In this paper we reconstruct $P_{in}(k)$ as a free function using
two different methods,
the wavelet band powers method of MW03a,
and the top-hat binning method of \cite{Wang99}.
We briefly describe our methods in Sec.2. Sec.3 contains our results.
Sec.4 contains a summary and discussions.

\section{Methods}

Both the methods that we have used to reconstruct $P_{in}(k)$ are 
essentially binning methods in which we can write $P_{in}(k)$ as
\be 
P_{in}(k) = \sum_i \alpha_i f_i(k),
\label{eq:Pink}
\ee
where $f_i(k)$ are functions of wavenumber $k$, and $\alpha_i$ are constants.
 We use Eq.(\ref{eq:Pink}), instead of  
 $P_{in}(k) = A\, k^{n_S-1}$, to parametrize $P_{in}(k)$
as an arbitrary function. 
Our $P_{in}(k)$ parameters are the coefficients $\alpha_i$'s, 
instead of the normalization parameter $A$,
the power-law spectral index, $n_S$, and its running, 
$\mbox{d} n_S/\mbox{d}\ln k$ \citep{Kosow95}.

This allows us to expand the CMB temperature angular 
power spectrum as follows:
\begin{eqnarray}
C_l(\{\alpha_i\}, \mbox{\bf s}) &=&(4\pi)^2 \int \frac{dk}{k} P_{in}(k) 
\left|\Delta_{Tl}(k, \tau=\tau_0)\right|^2
\nonumber\\
&=& \sum_{i} \alpha_i \int \frac{dk}{k} f_i(k)
\left|\Delta_{Tl}(k, \tau=\tau_0)\right|^2\nonumber\\
&\equiv & \sum_i \alpha_i \, C_l^i (\mbox{\bf s}),
\label{Clwaveletproj}  
\end{eqnarray}
where the cosmological model dependent transfer function 
$\Delta_{Tl}(k,\tau=\tau_0)$ is an integral over
 conformal time $\tau$ of the sources which generate CMB 
 temperature fluctuations,
 $\tau_0$ being the conformal time today,
 and $\mbox{\bf s}$ represents cosmological parameters other than
the $\alpha_i$'s. We use CAMB\footnote{Similar to
CMBFAST, CAMB can be used to compute the CMB and 
matter power spectra from a given set of cosmological parameters and primordial 
power spectrum. For details, see http://camb.info/.} to compute the 
CMB angular power spectra,
 in a form such that for given cosmological 
parameters other than the $\alpha_i$'s, the $C_l^i(\mbox{\bf s})$ are computed,
so that there is no need to call CAMB when we vary only the $\alpha_i$'s.

The choice of the basis functions $f_i(k)$ in Eq.(\ref{eq:Pink}) 
differ in the two methods as described below.

\subsection{The wavelet band powers method}

In this method, we are using wavelets essentially as band pass filters.
The wavelet band powers of the primordial power spectrum are given by
(MW03a)
\begin{equation}
P_j = \frac{1}{2^j} \sum_{n=-\infty}^{\infty} \left| \hat{\psi}
\left(\frac{n}{2^j}\right) \right|^2 P_{in}(k_n).
\label{impeqn1}
\end{equation}
The wavelet band power window functions in $k$ space, 
$\left| \hat{\psi}\left(\frac{k}{2^j}\right)\right|^2$,
are the modulus squared 
of the Fourier transforms of the wavelet basis functions of different $j$ 
(dilation index). The translation index has been integrated out as we  
Fourier transformed the basis functions. The resulting band powers $P_j$'s 
are thus band averaged Fourier power spectrum. 

The wavelet band power window functions, 
$\left| \hat{\psi}\left(\frac{k}{2^j}\right)\right|^2$, 
are plotted in Fig.1 of MW03a. 
Fig.2 of MW03a shows the window functions in CMB
multipole $l$ space that these functions map on to.

If the primordial density field is a Gaussian random field, the 
$P_j$'s, which represent the variance of wavelet coefficients of scale $j$,
 are uncorrelated (e.g. \cite{pia00}):
\be
\frac{ \langle P_j\, P_{j'} \rangle} {P_j \,  P_{j'}}
= 1.
\ee

Furthermore, the primordial power spectrum $P_{in}(k)$ can be reconstructed as 
a smooth function from
the wavelet band powers $P_j$'s as follows (\cite{Fang00}, MW03a)
\begin{equation}
\hat{P}_{in}(k) = \sum_{j} P_j \left| \hat{\psi}\left(\frac{k}{2^j}
\right)\right|^2,
\label{impeqn2}
\end{equation}
i.e., $\alpha_i=P_i$, and $f_i(k)=\left| \hat{\psi}\left(\frac{k}{2^i}
\right)\right|^2$ in Eq.(1).
MW03a has shown that Eq.(\ref{impeqn2}) gives excellent estimates
of $P_{in}(k)$ at the centers of the wavelet window functions. 
Smooth wavelets work best in this method. We have chosen 
the wavelet Daubachies 20 \citep{Daub92}. Our results are insensitive
to the choice of the particular wavelet among smooth wavelets.

The wavelet band powers method is an optimal binning method,
in which the locations of the bands is not arbitrary. 
Here the position
and momentum spaces are decomposed into elements that satisfy
$\Delta x\, \Delta k \sim 1$, with $\Delta x \propto 1/k$,
and $\Delta k/k = \log_{10}\,2$. Thus on small length scales
(large $k$), $\Delta x$ is small, and on large length scales,
$\Delta x$ is large, and since the wavelet bases are complete,
 one cannot have more independent bands than used here \citep{Fang00}.
We choose to estimate 11 $P_j$'s 
that cover the $k$ range probed by current data. The $P_j$'s outside
of this $k$ range are set equal to their adjacent $P_j$'s.

Note that although the $P_j$'s are mutually uncorrelated by construction,
the $P_j$'s estimated from CMB data (the
measured CMB temperature anisotropy angular power spectrum $C_l$ bands) 
will be somewhat correlated
because of the cosmological model dependent nonlinear mapping between the wavenumber $k$ and the 
CMB multipole number $l$, and due to correlations with 
cosmological parameters.

 We have chosen to use the wavelet band powers method in this paper,
since the first year WMAP data seem to be relatively well fit by a 
smooth $P_{in}(k)$ and are not yet sensitive to very sharp features
(see Fig. 7 of \cite{Peiris03}). The direct wavelet expansion method
 of \cite{pia2} is more suited to reconstructing an unknown function with 
possible sharp features on scales smaller than $log_{10}$$\,2$.

\subsection{The top-hat binning method}

We also use the top-hat binning method of \cite{Wang99} to  
parametrize $P_{in}(k)$. We write
\be
P_{in}(k)  = \left\{ \begin{array}{lll} 
\alpha_1, \hskip 1cm k < k_{0}; \\
\alpha_i, \hskip 1cm k_{i-1}< k < k_i; \\
\alpha_{n}, \hskip 1cm k > k_{n}.
\end{array} \right.
\ee
The $k_i$'s are chosen to be uniformly spaced in $\log\,k$ as in
\cite{Wang99}. 
In this method, the basis functions $f_i(k)$ in 
Eq.(\ref{eq:Pink}) are top-hat window functions: 
$f_i(k) = 1$, for $ k_{i-1} < k < k_i$, and $f_i(k) = 0$ elsewhere.
The boundary conditions at the minimum and maximum $k$ values
are similar to what we imposed in the wavelet band powers method.

We choose the centers of the top-hat window functions to 
coincide with the central $k$ values of the wavelet band power window
 functions, so that the results from the two methods can be compared
 for consistency.

The advantage of this binning is its simplicity.
The disadvantage is that the reconstructed $P_{in}(k)$
is a {\it discontinuous} step function, which might introduce
additional degeneracies with cosmological parameters.

We have included this binning method primarily to cross-check
the results of the wavelet band powers method.

\section{Results}

We work with CMB temperature anisotropy data from WMAP 
\citep{Bennett03}, complemented at $l>800$, and upto an $l_{max}$ of 2000,
 by data from 
CBI \citep{Pearson02} and 
ACBAR \citep{Kuo02}, and large scale 
structure (LSS) power spectrum data from 
the 2dFGRS \citep{Percival02} and PSCZ \citep{Hamilton02} 
galaxy redshift surveys.
We use the data covariance matrices and window functions provided by 
the different experimental teams. For CMB data we marginalize analytically 
over known beanwidth and calibration uncertainties (Bridle et al. 2002). 
For LSS data we assume that the galaxy power spectrum
 is a multiple of the underlying matter power spectrum and marginalize
 analytically over a linear bias \citep{Lewis02}. Both CMB and LSS data
depend on $P_{in}(k)$ and on the cosmological parameters.
Note however that as pointed out by \cite{Elgaroy02}, it is hard to detect 
features in the primordial power spectrum at $k<0.03 h$Mpc$^{-1}$ 
using LSS data. 

We estimate the $P_{in}(k)$ parameters $\alpha_i$'s (see
Eq.(\ref{eq:Pink})), together with 
the Hubble constant $H_0$, baryon density $\Omega_b\,h^2$, 
cold dark matter density $\Omega_c\,h^2$, and reionization 
optical depth $\tau_{ri}$.
We assume Gaussian adiabatic scalar perturbations in a flat universe
 with a cosmological constant. We do not use tensor modes in this paper,
 since current data are not sensitive to tensor contributions. 
 We make use of the WMAP constraint on $\tau_{ri}$, derived
for a $\Lambda$CDM cosmology from WMAP's TE polarization data 
\citep{Kogut03}, by
imposing a Gaussian prior on $\tau_{ri}$, $p(\tau_{ri}) \propto
\exp\left[ - (\tau_{ri}-0.17)^2/(2 \sigma_{\tau_{ri}}^2)\right]$,
with $\sigma_{\tau_{ri}}=0.04$ (this error estimate includes systematic
 and foreground uncertainties).

For a power-law primordial fluctuation spectrum,
$P_{\cal R}= 2.95 \times 10^{-9} \, A\,(k/k_0)^{n_S-1}$
\citep{Spergel03}. For an arbitrary primordial power spectrum,
we define 
\be
P_{in}(k)=\frac{P_{\cal R}}{2.95 \times 10^{-9}}.
\ee
 We use $k_0 = 0.05\,$Mpc$^{-1}$ when quoting 
parameter constraints for a power law model.

We use the Markov Chain Monte Carlo (MCMC) technique to estimate
 the likelihood functions of the parameters
\citep{neil,Knox01,Kosowsky02,Lewis02,Verde03}. 
The use of MCMC is necessitated by the large number of parameters,
and it is free of the interpolation errors expected in the conventional
and much slower grid method.
At its best, the MCMC method scales approximately linearly with the number of 
parameters. The method samples from the full posterior distribution of the
 parameters, and from these samples the marginalized posterior distributions
 of the parameters can be estimated.  

Table 1 lists the mean values and marginalized 1 $\sigma$ confidence limits 
of the cosmological parameters estimated using four different models
for $P_{in}(k)$, as well as $\chi^2_{eff}= -2 \ln {\cal L}$ where ${\cal L}$ is
the likelihood of each model.
Both the wavelet bandpowers model and the top-hat binning model allow
$P_{in}(k)$ to be an arbitrary non-negative function.
The scale invariant model assumes $P_{in}(k)=A$ ($A$ is a constant),
while the powerlaw model assumes $P_{in}(k)$ to be a powerlaw,
$P_{in}(k)=A k^{n_S-1}$ ($A$ and $n_S$ are constants). 
The only priors used 
are a Gaussian prior on $\tau_{ri}$, $p(\tau_{ri}) \propto
\exp\left[ - (\tau_{ri}-0.17)^2/(2 \sigma_{\tau_{ri}}^2)\right]$,
with $\sigma_{\tau_{ri}}=0.04$ as discussed earlier, and a weak age 
prior of age of the universe $t_0 > 10$ Gyrs.
Also, we do not use 
tensor modes 
in this paper,
since current data are not sensitive to tensor contributions.

Fig.1 shows the reconstructed $P_{in}(k)$ from the two different
methods discussed in Sec.2, using CMB temperature anisotropy data.
The dotted line in each panel indicates the scale-invariant model that fits
 these data\footnote{The uncertainty in $P_{in}(k)$ in the wavelet band
 powers method is calculated using $\sigma^2_{P_{in}(k)} = \frac{1}{N}
 \sum_N \left [ \sum_j (P_j-\bar{P}_j) \left| \hat{\psi}
\left(\frac{k}{2^j}\right) \right|^2 \right ] ^2$, where the average
 is over the MCMC samples and $\bar{P}_j$ denotes the mean or expectation
 value of the wavelet band power.} .

From Table 1 we see that the power-law model differs by 
$\Delta\chi^2_{eff}=8$ from the
the wavelet band powers model, and by $\Delta\chi^2_{eff}=10$
from the top-hat binning model. Since the difference in the 
number of degrees of
freedom is approximately 9, the power-law model is disfavored
at approximately $\sim 0.7\,\sigma$ and $\sim 1\, \sigma$
compared to the wavelet band powers model and
the top-hat binning model respectively.
Also, note that the powerlaw model is favored over the 
scale-invariant model at less than $1\,\sigma$.

Given the estimated wavelet band powers $P_j$'s with their full covariance matrix
$C_{P_j}$,
we can treat the $P_j$'s as data
and compute $\chi^2 \equiv \Delta^T C_{P_j}^{-1} \Delta$, 
$\Delta$ being the difference between the estimated $P_j$'s 
 and the $P_j$'s corresponding to the fitted power-law and 
scale-invariant spectra. 
 The $\chi^2$'s turn out to be 8 and 7 
 for the power-law and scale-invariant parametrization
 respectively. In the top-hat binning case, we can treat the estimated
  top-hat bin amplitudes as data
 and compute a similarly defined $\chi^2$.
 We find that the corresponding $\chi^2$ is 9 for both power-law and scale-invariant 
 parametrizations.
 Thus similar significances for
deviation of the reconstructed $P_{in}(k)$
 from the simpler parametrizations are indicated in this way also.
 In general, note that the low
 levels of significance are also due to the large number of degrees of
 freedom that we are allowing for in the analysis here, and it is clear 
from the figure which points do not contribute much to the $\chi^2$.

%\pspicture(0,0.2)(5.5,12.4)
%\rput[tl]{0}(-0.2,12.2){\epsfxsize=8.5cm \epsfclipon
%\epsffile{f1.eps}}
%\rput[tl]{0}(0,3.3){
%\begin{minipage}{8cm}
%\small\parindent=3.5mm
%{\sc Fig.}~1.---

\clearpage

\begin{figure}
\plotone{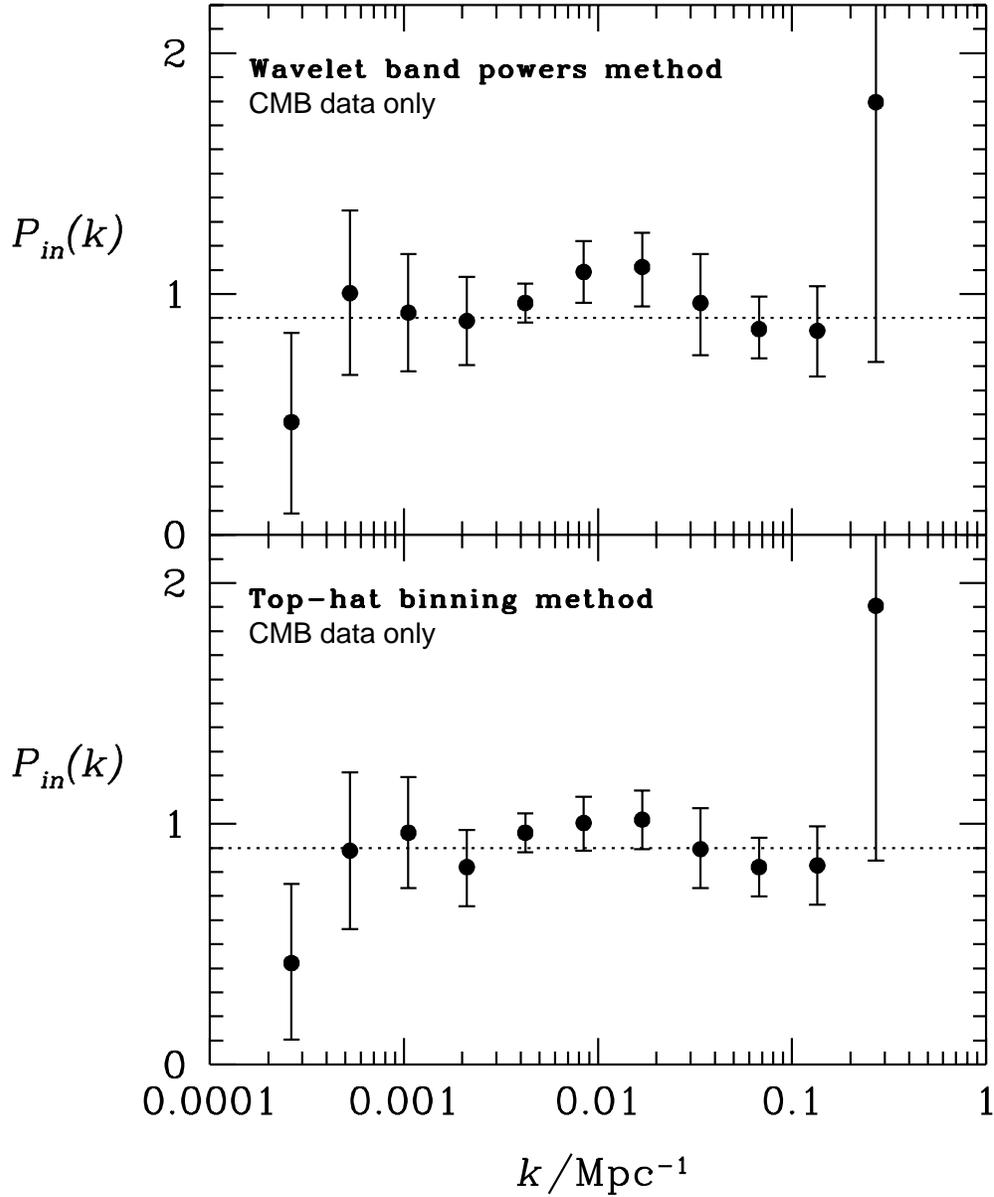}
\figcaption{
the reconstructed $P_{in}(k)$ with 1$\sigma$ error bars from the two different
methods discussed in Sec.2, using only CMB data.
The dotted line indicates the scale-invariant model that best fits the data.
}
\end{figure}

%\end{minipage}
% }
% \endpspicture
% \vskip -1cm
 
\clearpage

Fig.2 shows the reconstructed $P_{in}(k)$ from the two different
methods discussed in Sec.2, using CMB temperature anisotropy data 
as above together with 
LSS data from the 2dFGRS and PSCZ galaxy redshift surveys.
The constraints on cosmological parameters are listed in Table 1.

The results of fitting the same CMB and LSS data to a scale-invariant model
and a power-law model are shown in Table 1.
The power-law model differs by 
$\Delta\chi^2_{eff}=6$ from the
the wavelet band powers model\footnote{Since the two
methods give very similar estimates of $P_{in}(k)$, 
the smaller $\chi^2_{eff}$ of the top-hat binning method
seems to indicate that the wavelet band power method 
has not yet sampled the parameter values 
with the smallest possible $\chi^2_{eff}$.
}, and by $\Delta\chi^2_{eff}=9$
from the top-hat binning model. Since the difference in the 
number of degrees of
freedom is approximately 9, the power-law model is disfavored
at approximately $\sim 0.4\,\sigma$
 and $\sim 0.9\, \sigma$
compared to the wavelet band powers model and
the the top-hat binning model respectively.
The powerlaw model is again favored over the 
scale-invariant model at less than $1\,\sigma$.

As discussed previously, the estimated parameters that describe the
power spectrum as an arbitrary function (wavelet band powers or
the tophat bin amplitudes) can be treated as data,
and compared with the fitted power-law and scale-invariant
spectra
by computing a $\chi^2$ using the full covariance matrix of
the estimated parameters (wavelet band powers or
the tophat bin amplitudes).
In the wavelet band powers method,
the $\chi^2$ for both power-law and scale-invariant parametrizations is 6.
Similarly in the top-hat case the $\chi^2$'s 
are 10 and 12 for the power-law and scale-invariant parametrization
respectively.
These indicate similar significances for deviation of the 
reconstructed spectrum from the simpler parametrizations.

%\pspicture(0,0.2)(5.5,12.4)
%\rput[tl]{0}(-0.2,12.2){\epsfxsize=8.5cm \epsfclipon
%\epsffile{f2.eps}}
%\rput[tl]{0}(0,3.3){
%\begin{minipage}{8.75cm}
%\small\parindent=3.5mm
%{\sc Fig.}~2.---

\clearpage

\begin{figure}
\plotone{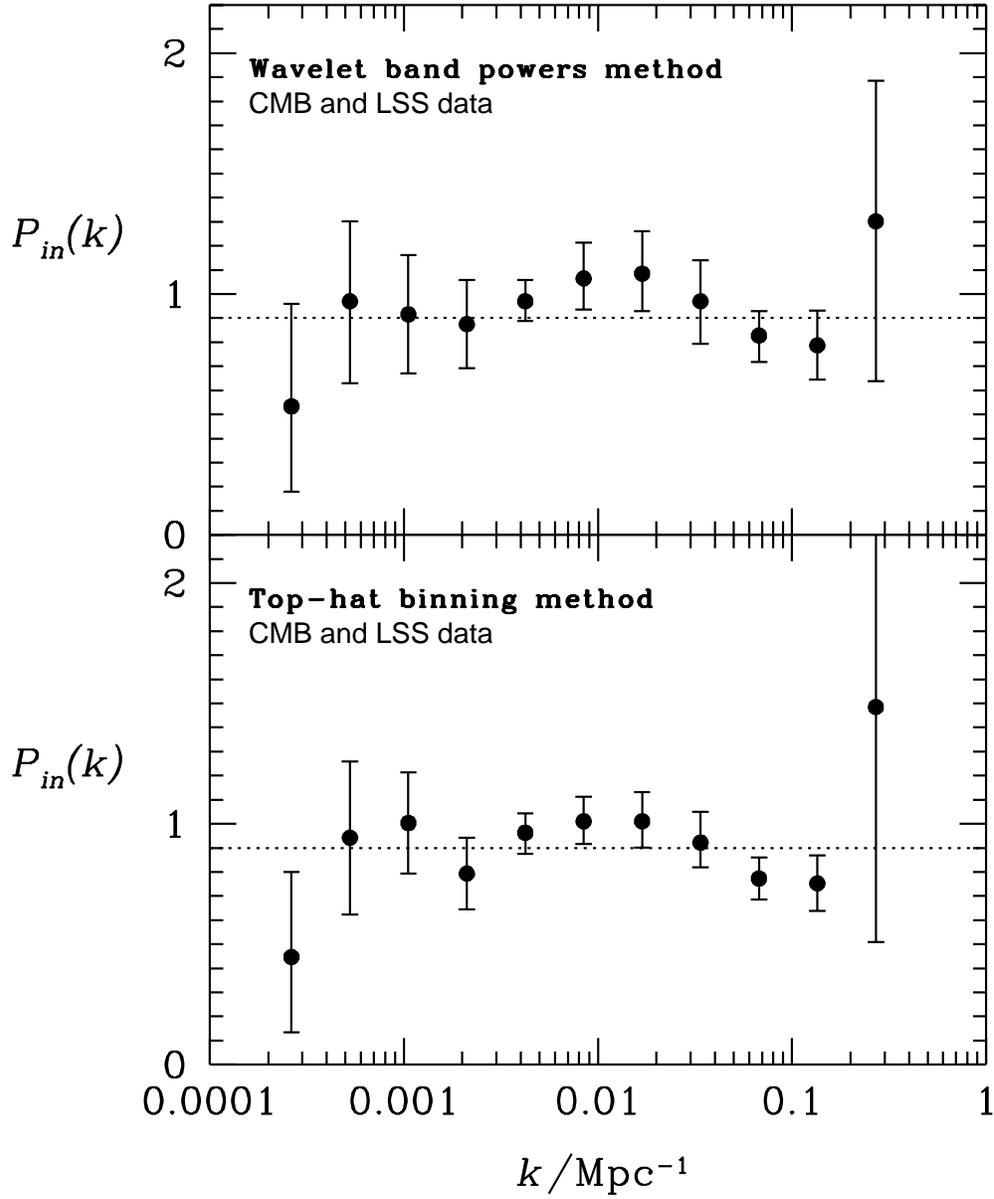}
\figcaption{
The reconstructed $P_{in}(k)$ with 1$\sigma$ error bars from the two different
methods discussed in Sec.2, using CMB data and
LSS data.
The dotted line indicates the scale-invariant model that best fits the data.}
\end{figure}

\clearpage

%\end{minipage}
%}
% \endpspicture

%\vskip -1cm

Fig.3 shows the reconstructed $P_{in}(k)$ using the wavelet band powers
method, compared with the $P_{in}(k)$ constraints derived using 
the WMAP team's constraints on $A$, $n_S$ and 
$\mbox{d} n_S/\mbox{d}\ln k$ for $P_{in}(k)=A\,(k/k_0)^{n_S-1}$
\citep{Peiris03,Spergel03} (shaded region).\footnote{
The WMAP constraints in Fig.3 is similar to Fig.2 
of \cite{Peiris03} (which considers tensor contributions), but the 
$P_{in}(k)$ parameter constraints are taken from Table 8 of 
\cite{Spergel03}, since we do not consider 
tensor contributions in this paper.}
The shaded region in Fig.3 are only
meant to illustrate roughly the WMAP team's constraints, since we
have not included the covariances among $A$, $n_S$ and 
$\mbox{d} n_S/\mbox{d}\ln k$ estimated by them (these are not publicly 
available). Clearly, our results are consistent with the WMAP results within
1$\,\sigma$.

\clearpage

%\pspicture(0,0.2)(5.5,12.4)
%\rput[tl]{0}(-0.2,12.2){\epsfxsize=8.5cm \epsfclipon
%\epsffile{f3.eps}}
%\rput[tl]{0}(0,3.3){
%\begin{minipage}{8.75cm}
%\small\parindent=3.5mm
%{\sc Fig.}~3.---

\begin{figure}
\plotone{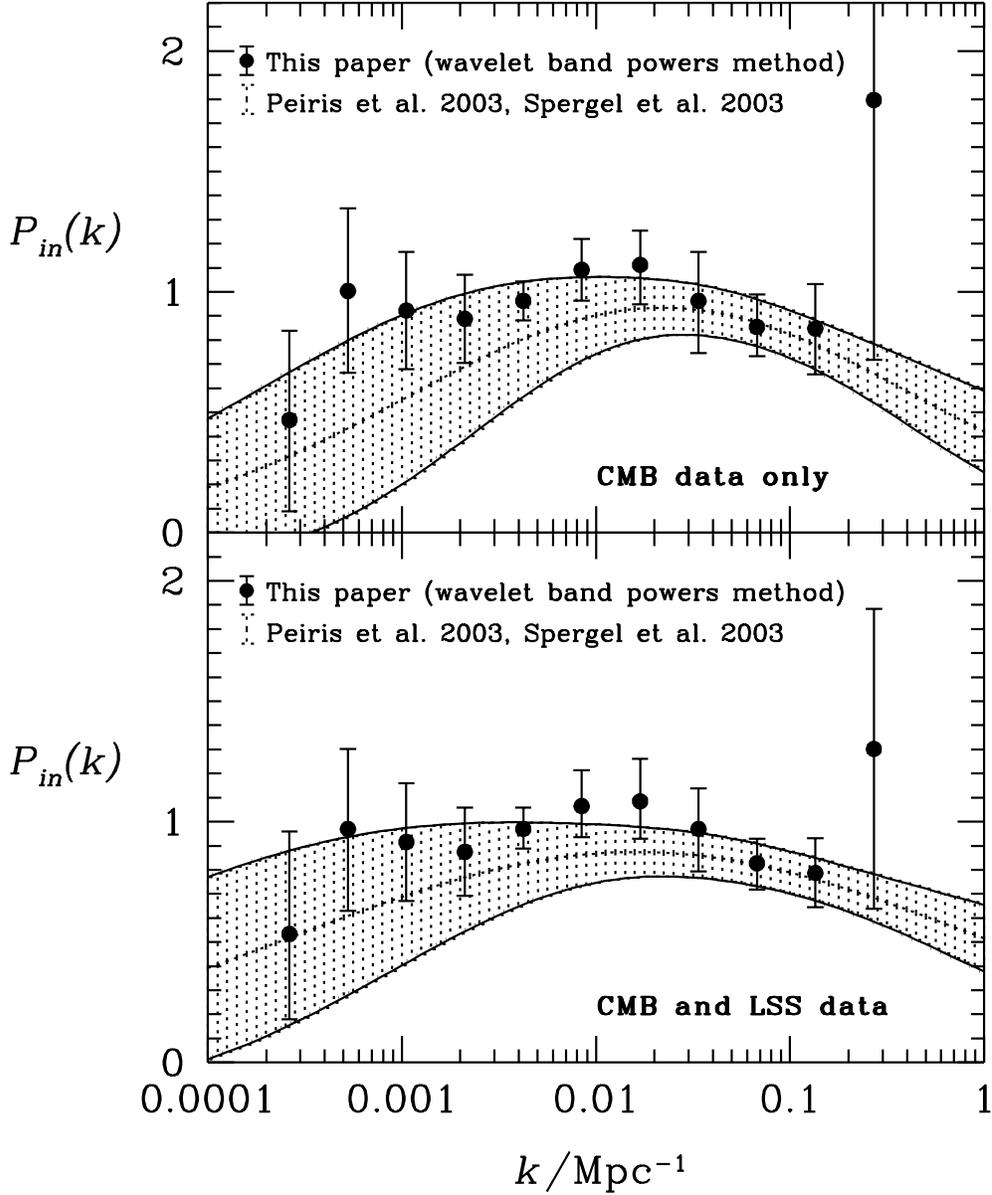}
\figcaption{
The reconstructed $P_{in}(k)$ using the wavelet bandpower
method (with 1$\sigma$ error bars), compared with the $P_{in}(k)$ constraints 
derived using the WMAP team's constraints on $A$, $n_S$ and 
$\mbox{d} n_S/\mbox{d}\ln k$ for $P_{in}(k)=A\,(k/k_0)^{n_S-1}$
\citep{Peiris03,Spergel03} (shaded region). We have not included the 
covariances among
$A$, $n_S$ and $\mbox{d} n_S/\mbox{d}\ln k$.
}
\end{figure}

\clearpage

%\end{minipage}
% }
%\endpspicture
% \vskip -1cm

We note that the cosmological parameters are relatively 
well constrained even when the primordial power spectrum is 
reconstructed as a free function. In the MCMC method, the 
parameters are allowed to vary within wide limits. We have monitored
 convergence and mixing as advocated in Verde et al. (2003).
 The amplitude of the band 
on the smallest scale (centered at $k\sim 0.2\,$Mpc$^{-1}$) is
essentially unconstrained in both the wavelet band power and top hat binning
 methods when using just CMB data. 
 Using CMB and LSS data,
 the amplitude of this band gets constrained. 
 The amplitude in the band centered at $k\sim 0.0003\,$Mpc$^{-1}$
 is single tailed towards larger values, but well constrained within the prior. 
 Besides these bands, the power in all the other
 bands and the cosmological parameters are all well constrained, and have
 close to Gaussian 1d marginalized distributions.  Parameter constraints
from the full $n$-D distribution are somewhat weaker,
 as expected, but consistent with the 1d marginalized distributions.

We find only slight evidence for a preferred scale at $k \sim 0.01\,$Mpc$^{-1}$
in the primordial power spectrum from current data (Figs.1-2).
This apparent deviation from scale-invariance of $P_{in}(k)$ accompanies
a slightly low Hubble constant, and non-vanishing 
reionization optical depth $\tau_{ri}$ (see Table 1).
Inclusion of tensor contributions in the analysis would also
 increase the effect 
(as the data would then be consistent with reduced power on large 
scales which can be filled in by tensor contributions, as also noted 
in \cite{Seljak03}). The data do not require this deviation however, and 
within parameter degeneracies appear consistent with scale invariance, 
as well as with a slight red tilt (see Table 1). 
Note that the current data are consistent with the tensor to scalar
ratio $T/S=0$, and the fit is not improved by including $T/S$ as a parameter
\citep{Spergel03}. Therefore, current data
do not require a tensor contribution. However, this 
does not imply that a non-zero tensor contribution 
is ruled out.
Similarly, deviation of the primordial power spectrum 
from scale invariance is not ruled out at present, though limits 
can be placed on such deviations (see Figures).  
We would be able to better distinguish between these models if
 cosmological parameters could be constrained to better accuracy.
We have not included Lyman $\alpha$,
 weak lensing and supernovae data, since these have larger uncertainties
 at present.

\section{Summary and Discussion}

Reconstructing the shape of the primordial power 
spectrum $P_{in}(k)$ in a model
independent way from cosmological data is a useful consistency check 
on what is usually assumed regarding early universe physics.  It is also
our primary window to unknown physics during inflation. We have used two 
methods to reconstruct
$P_{in}(k)$ as a free function from CMB temperature
anisotropy and LSS data (Figs.1 and 2). 
The two methods are complementary to each other, and 
give consistent results.
We find that $P_{in}(k)$ reconstructed from CMB data alone (WMAP, 
CBI, and ACBAR), or from CMB data together with LSS data 
(2dFGRS and
PCSZ), seems to
indicate excess power for $ 0.002\,$Mpc$^{-1} \la k \la 0.03\,$Mpc$^{-1}$,
consistent with that found by the WMAP team \citep{Peiris03}
but at a lower significance of $\sim 1\,\sigma$ (Fig.3).
Note that the significance level deduced here is also low because we 
are reconstructing $P_{in}(k)$ in a large number of bins.
Neither a scale-invariant $P_{in}(k)$ nor a power-law $P_{in}(k)$ 
is ruled out by the current data.

We find that this apparent deviation of $P_{in}(k)$ accompanies
a slightly low Hubble constant (and correspondingly 
a slightly high $\Omega_m$), and a non-vanishing $\tau_{ri}$ (Table 1).
However, the 1$\sigma$ error bars on our derived $H_0$
values
overlap with the 1$\sigma$ error bar obtained by the 
HST Key project \citep{Freedman01} and matches well the $H_0$
 determined using supernovae \citep{Branch98}. 
Note that because of parameter degeneracies, the
 $H_0$ values derived from CMB data represent {\it indirect} measurements,
 while the $H_0$ derived from Cepheid distances \citep{Freedman01}
 or supernova data \citep{Branch98} are {\it direct} measurements.
 It is also important to include the systematic uncertainty
 in local direct measurements of $H_0$ due to matter inhomogeneity
 in the universe \citep{Wang98}, as included in the error estimate
 of $H_0$ by \cite{Freedman01}.
Clearly, more stringent independent measurements of $H_0$ 
can help tighten the constraints on $P_{in}(k)$.
 
We have not included tensor contributions in our analysis, because
the WMAP data are not yet constraining on the tensor perturbations
from inflation. The inclusion of tensor contributions are expected
to increase the deviation of $P_{in}(k)$ from scale-invariance.

Our results are consistent with that of \cite{Bridle03} and \cite{Barger03}; 
both find that the $P_{in}(k)$ derived from current data from WMAP (\cite{Bridle03} 
included LSS data as well) are consistent with
scale-invariance. 
There are two basic differences between the analysis presented in \S4 of \cite{Bridle03} and this paper.
The sensitivity of their method to the 
feature around $k \sim 0.01$Mpc$^{-1}$ is reduced because their banding
 oversamples this region by a factor of three.
Also, \cite{Bridle03} reconstructed $P_{in}(k)$ using linear interpolation
of amplitudes of $P_{in}(k)$ at discrete $k$ points, an approach 
pursued in \cite{WangMathews02} and MW03a. This method is expected to lead to 
stronger correlations between adjacent $P_{in}(k)$ amplitudes estimated 
from data. Since the different binning choices made by us and \cite{Bridle03}
lead to different correlations between the 
estimated parameters, they provide complementary and somewhat
different information. \cite{Barger03} used WMAP temperature data to constrain slow-roll 
inflationary models. 
%The difference in results between \cite{Barger03} 
%and this paper can be explained as follows.
They find that $\tau_{ri}=0$ is preferred based on
WMAP temperature data. 
We have found that taking $\tau_{ri}=0$
greatly diminishes the significance of any deviations of $P_{in}(k)$
from scale-invariance.
We have chosen to take into consideration the implications of the
WMAP polarization data 
by applying a Gaussian prior on $\tau_{ri}$ 
based on the results of \cite{Kogut03}.

We note that Miller et al. (2002) have
 examined the pre-WMAP CMB temperature data in a non-parametric way to check
 whether the data can be better fit by breaking away from our assumed 
cosmological model, and to deduce the 
significance levels of the acoustic peaks in the $C_l$ spectrum 
non-parametrically. This also helps test the robustness of the 
cosmological model. 

We conclude that without making assumptions about the form of $P_{in}(k)$,
the $P_{in}(k)$ derived from first year WMAP data deviates from
 scale-invariance (with a preferred scale at 
$k \sim 0.01\,$Mpc$^{-1}$) only at a significance level
of approximately $1\,\sigma$ (Fig.1-2).
Simplest forms of $P_{in}(k)$ (scale-invarant, or power-law) 
are thus consistent with the data at present.
The WMAP data in subsequent years, together with improved constraints
from other independent cosmological probes, will allow us to place firmer
constraints on very early universe physics.

\acknowledgements
It is a pleasure for us to thank Hiranya Peiris and Dipak Munshi
for helpful discussion,
the referee for useful comments, 
and Henry Neeman for computational assistance.
We acknowledge the use of CAMB and CosmoMC. This work is
supported in part by NSF CAREER grant AST-0094335.

\begin{deluxetable}{lccccccccc}
\tablecolumns{8} 
\tabletypesize{\scriptsize\tiny}
%\tablewidth{0pt}
\tablecaption{Parameters Estimated from CMB\tablenotemark{a} and 
LSS\tablenotemark{b} data\tablenotemark{c}}
\tablehead{
$P_{in}(k)$ model & data used & $P_{in}(k)$ parameters &
$\Omega_b\,h^2$ & $\Omega_m\,h^2$
& $h $ & $\tau_{ri}$ & $\chi^2_{eff}$\\ 
}
\startdata
wavelet band powers    &  CMB only  & see Fig.1 &
 $.0180  \pm .0038$ &
$.143 \pm  .029$ & $ .575 \pm   .082$ & $.185 \pm  .045$ & 980.04 \\
top-hat binning    &  CMB only  & see Fig.1 &
$.0185 \pm  .0031$ &
$.129 \pm .029$ & $ .617 \pm  .080$  & $ .175 \pm  .044$ & 977.81\\
scale-invariant    &  CMB only  & $A=.893 \pm .050 $ &
$.0237 \pm .0006$ & $.123 \pm .015$
 & $.710\pm  .044 $& $.173\pm .036$   & 988.29\\
power-law    &  CMB only  & $A=.799\pm .117$ & 
 $.0228\pm .0012$& 
 $.116\pm .016$  & $.713\pm .044 $& $.136\pm .054$ & 987.92\\
 & & $n_S=.974\pm .028$ & & & & & \\
\hline
wavelet band powers    &  CMB \& LSS  & see Fig.2 &
$.0187\pm .0031$ &
$.136  \pm .021$ & $ .601 \pm  .069 $ & $.191 \pm .047$ & 1037.86 \\
top-hat binning    &  CMB \& LSS  & see Fig.2 &
$.0189 \pm  .0019$ &
$.134 \pm  .016$ & $ .597 \pm  .049 $ & $ .164 \pm  .047 $ & 1034.80\\
scale-invariant    &  CMB \& LSS  & $A=.883\pm .050$  &
$.0238\pm .0006$ & 
$.121\pm .007$  & $.714\pm .022$ & $.170\pm .032$ & 1044.33\\
power-law    &  CMB \& LSS  & $A=.836\pm .107$ & 
 $.0233 \pm .0010$   & 
 $.120\pm .007$ & $.707\pm .026$ & $.147\pm .055$ & 1043.68\\
 & & $n_S=.985\pm .028$ & & & & & \\
\enddata
\tablenotetext{a}{CMB 
temperature anisotropy data from WMAP, CBI and ACBAR}
\tablenotetext{b}{LSS power spectrum data from the 2dFGRS and PSCZ galaxy redshift surveys}
\tablenotetext{c}{The number of data points in the different data sets used are 899 (WMAP), 4 (CBI), 7 (ACBAR), 32 (2dFGRS) and 22 (PSCZ).} 
\end{deluxetable}

\end{document}